# Impact of disorder on the 5/2 fractional quantum Hall state

W. Pan[1], N. Masuhara[2], N.S. Sullivan[2], K.W. Baldwin[3], K.W. West[3], L.N. Pfeiffer[3], and D.C. Tsui[3]

[1] Sandia National Labs, Albuquerque, New Mexico 87185, USA

[2] University of Florida and National High Magnetic Field Laboratory, Gainesville, Florida 32611, USA

[3] Princeton University, Princeton, New Jersey 08544, USA

Abstract:
We compare the energy gap of the $\nu$=5/2 fractional quantum Hall effect state obtained in conventional high mobility modulation doped quantum well samples with those obtained in high quality GaAs transistors (heterojunction insulated gate field-effect transistors). We are able to identify the different roles that long range and short range disorders play in the 5/2 state and observe that the long range potential fluctuations are more detrimental to the strength of the 5/2 state than short-range potential disorder.

The interplay between electron-electron (e-e) interaction and disorder plays an important role in determining many-body ground states in low-dimensional systems. One example is the apparent metal-to-insulator transition in two-dimensional electron systems (2DES) [1]. From the scaling theory by Abrahams, *et al* in 1979 [2], disorder destroys any 2D extend states and, in the absence of e-e interaction, renders the system an insulator at T=0. The discovery of an apparent metallic state in the strongly interacting regime in high quality 2DES in the early 90's [1] highlights the complex interplay between e-e interaction and disorder. Another area where this interplay has a profound impact is the fractional quantum Hall effect (FQHE) [3,4]. Since the first discovery of the $\nu$=1/3 fractional quantum Hall effect (FQHE) state, experimental and theoretical evidence has accumulated that sample disorder plays an crucial role in the stability of the FQHE ground states. A FQHE state is only observed in specimens in which the 2DES mobility (related to sample disorder) exceeds a certain threshold [5,6].

Among all observed fractional quantum Hall effect (FQHE) states, the $\nu$=5/2 FQHE state [7,8] in the second Landau level remains the most exotic one. This state has been at the center of current quantum Hall research due to the possibility of it being non-Abelian [9,10] and, thus, having potential applications in fault-tolerant topological quantum computation [11].Therefore, there is an urgent need to understand the impact of disorder on this FQHE state, since a larger energy gap at $\nu$=5/2 would exponentially reduce error rates [11] and make the envisioned quantum computation more robust.

Today, our knowledge of the impact of various kinds of disorder on the 5/2 FQHE, or any other of the FQHE states, remains limited. Indeed, it has long been observed [12,13] that in the lowest Landau level, even after taking into account the finite thickness correction of 2DES and Landau level mixing effect, there is still a discrepancy of ~ 2-3K between the experimentally measured energy gaps of the odd-denominator FQHE states and theoretically calculated ones. This value of discrepancy has also been observed for the 5/2 state and the odd-denominator FQHE stats in the second Landau level [8]. In order to reconcile the discrepancy, often, a very vague quantity termed "disorder broadening" is employed [8], which does not seem to correlate with the 2DES mobility. Moreover, little is known about the nature of disorder, e.g., the different influence of long range Coulombic and short range neutral disorders, on the energy gap.

On the other hand, understanding the impact of disorder is essential for understanding the nature of a quantum system. As shown in a series of papers [14], the long standing controversies on the universality of quantum Hall plateau transition is directly linked to the nature of disorder in the 2D channel. In 2DES samples with strong disorder from a short-range neutral alloy potential a perfect power law scaling behavior of quantum Hall plateau-to-plateau transition was observed over more than two decades of temperatures, from 1.2K down to 10 mK. In contrast, in samples with weak disorder from a long-range Coulombic potential a crossover behavior was observed from a non-universal scaling region at high temperatures to a universal scaling region at low temperatures.

In this letter, we present our results on how the nature of disorder affects the 5/2 energy gap and, thus, the stability of this state. We compare the activation energy gap data obtained in two types of samples: symmetrically doped modulation quantum well samples and undoped heterojunction insulate-gated field-effect transistors (HIGFETs). In modulation doped quantum well samples, where long-range Coulombic disorder dominates, the energy gap drops quickly with decreasing mobility (or increasing disorder). On the other hand, in HIGFET samples, where the short-range neutral disorder dominates, the 5/2 energy gap shows only a weak mobility dependence. Our results clearly demonstrate that the two types of disorder play very different roles in affecting the stability of the 5/2 state. Possible physical mechanisms are discussed.

For this study we used two HIGFET (heterojunction insulate-gated field-effect transistor) [15] devices, called A and B. Results from these two samples are consistent with each other. Figure 1 shows the mobility ($\mu$) versus electron density (n) obtained in HIGFET A. The mobility first increases with increasing n, reaches a peak of $\mu$ ~ $14 \times 10^6$ cm$^2$/Vs around n ~ $1.8 \times 10^{11}$ cm$^{-2}$, and then decreases with increasing n. The decrease of mobility with increasing density beyond n ~ $1.8 \times 10^{11}$ cm$^{-2}$ is due to surface roughness scattering [16], which is of short-range. The weak kink at n ~ $3.5 \times 10^{11}$ cm$^{-2}$ is probably due to the onset of occupation of the second electrical subband.

We studied the $\nu$=5/2 state at various densities. In Figure 2a, we show the data obtained in HIGFET B, taken at the high B/T facilities of National High Magnetic Field Laboratory at the University of Florida. Different from the conventional magneto-transport measurement, where the 2D electron density is fixed and the magnetic field is swept, here, the magnetic field is fixed and the gate voltage and hence the carrier densities is swept. This process minimizes heating from eddy currents. As shown in Figure 2a, a well developed 5/2 state is observed and its diagonal resistance ($R_{xx}$) minimum reaches a very low value of ~ 10 ohms at a density of n=$2.85 \times 10^{11}$ cm$^{-2}$. With decreasing electron density, $R_{xx}$ at $\nu$=5/2 increases, but the minimum remains visible down to n ~ $0.55 \times 10^{11}$ cm$^{-2}$.

The density dependence of the 7/3 state is similar to that of the 5/2 state. However, the 8/3 state shows a slightly different density dependence. Its strength first decreases with decreasing density and then seems to saturate at lower densities.

Figure 2b shows the $R_{xx}$ traces at a few selected temperatures taken at a magnetic field B = 4.68T. Clearly, the $R_{xx}$ minimum shows an activated behavior, i.e., $R_{xx}$ increases with increasing temperature. This is different from our previous study [17] and thus allows us to obtain an activation energy value for the 5/2 state, as shown in Fig.2c.

In Fig. 2d, we show the density dependence of the 5/2 energy gap in HIGFET A, in the density range where surface roughness dominates. In this density range, the 5/2 energy gap increases with increasing 2D electron densities. Two important features are apparent: 1) the increase of the 5/2 energy gap is not caused by an increase in electron mobility. The energy gap at $\nu$=5/2 increases by a factor of more than 3 from ~0.07 K at n ~1.8x10$^{11}$ cm$^{-2}$ to 0.24 K at ~4x10$^{11}$ cm$^{-2}$, while over the same density range the mobility actually decreases by a factor of 2.  2) the dependence of the energy gap on density is smooth and no sharp features are apparent.

To better appreciate the impact of disorder on the 5/2 state, we plot in Figure 3a the normalized 5/2 energy gap versus electron mobility (measured at zero magnetic field) for HIGFET A. The normalized energy gap, defined by $\Delta_{\nu=5/2}/e^2/\varepsilon l_B$, is used to eliminate the density dependence of the energy gap in different samples. For comparison, we include in Fig.3a our 5/2 energy gap data from high quality, symmetrical modulation-doped 30nm quantum wells reported in Ref. [18]. Due to the wide well width, the disorder in such quantum wells is expected to be dominantly by distant ionized impurities and hence long-ranged. We further include the results by other groups [19-24], again in modulation doped quantum wells with the same or even larger well width in which the disorder is expected to be also long-ranged. Only the high density data points in Ref. [23] are used. In Fig. 3b, the data of Fig. 2 are plotted as a function of 1/$\mu$, which is a rough measure of sample disorder. It is clear that the two types of samples, HIGFETs and quantum-wells, show very different mobility dependences. For the modulation doped samples, where sample disorder is of long range, the normalized energy gap decreases sharply with decreasing mobility or increasing disorder. An apparent mobility threshold for a non-zero 5/2 energy gap of ~10x10$^6$ cm$^2$/Vs, is obtained from the extrapolation of a linear fit. On the other hand, in the HIGFET where the disorder is dominantly caused by charge-neutral surface roughness scattering, which is of short-range, a weak mobility dependence is seen. In fact, the normalized energy gap increases slightly with decreasing mobility or increasing disorder. This contrasting behavior suggests that the long-ranged Coulombic and short-ranged charge-neutral disorders play very different roles in affecting the 5/2 energy gap.

Before we discuss possible mechanisms that may be responsible for the observed difference, let us first look at the finite thickness effect of 2DES. We calculated the thickness of the 2DES using the elf-consistent method for the symmetric quantum wells and the Fang-Howard wavefunction for the HIGFETs. For the majority of the QWs quoted in our paper, the effective thickness, define by w/$l_B$, is roughly the same, w/$l_B$ ~ 2. Here w is the finite thickness of the 2DES in the growth direction and $l_B$ the magnetic length at $\nu$=5/2. For the HIGFETs, due to the fact that w scales as n$^{-1/3}$ and $l_B \propto$ n$^{-1/2}$, w/$l_B$ $\propto$ n$^{-1/6}$ shows a much weaker density dependence. In our studied density range, it varies between 0.9 and 1.1. We notice that in a recent paper by Papić et al [25] the change in the overlap between the exact wavefaction and the Pfaffian model wavefunction is smooth and the difference is small as w/$l_B$ varies from 1 to 2. Thus, it is hard to imagine that this factor of 2 difference in our samples can present such a striking difference in the disorder dependence observed in our experiment.

In the following, we discuss several possibilities that may shad light on our understanding of the origin of the observed difference in the mobility dependence. One possibility is that the attractive Coulombic interaction between ionized donors in the doping layer and 2D electrons may affect the pairing of composite fermions [26,27] at $\nu$=5/2. Indeed, it is known that Coulomb interaction can cause fluctuations of the phase of the superconducting order parameter and, therefore, destroy superconductivity [28]. Recent studies by Umansky et al [29] and Gamez et al [30] also showed the importance of ionized dopants in suppressing the development of the 5/2 state as well as other quantum Hall states in the second and higher Landau levels.

A second possibility might be related to the size of the quasi-particles at $\nu$=5/2. In a recent study by Nübler et al [23], it was shown that the size of 5/2 quasiparticles is fairly large, ~ 10 magnetic length, or ~ 0.1 $\mu$m or larger in our density range. Such large size quasiparticles may not be affected at all by the nanometer or sub-nano meter size fluctuations from surface roughness, whereas the long-range fluctuations from remote impurities might assembles the quasi particles into poorly connected puddles of micron size, which affect macroscopic transport and hence the energy gap data.

We also want to address an earlier discussion [31] on long-range and short-range disorder potentials in affecting the bounded magneto-rotons [32,33] and the current carrying quasiparticles near impurities. It was argued [31] that around impurities the ground state was slightly deformed. Such a deformation is energetically less costly if rotons are being excited, which in turn affect the energy gap of a FQHE state. It is possible that in the presence of long range disorder, such deformations in the 5/2 ground state occurs around the perimeter of electron puddles. As a result, its energy gap is expected to depend on the puddle size, which, in long-range potentials such as the one created by remote impurities of quantum-well samples, is related to electron mobility. In HIGFET samples, on the other hand, the mobility is dominated by short range potential fluctuations from surface roughness and this disorder configuration is largely fixed after sample growth. Therefore, in this case, a weak mobility dependence on the 5/2 energy gap as the 2DES density is varied, is not unexpected.

Finally, our data also can shed some light on the controversy between the 5/2 state being spin polarized or spin-unpolarized. The red line in Fig.2d shows our best fit assuming a fully polarized ground state at $\nu$=5/2, with $\Delta$ given by $\Delta = \alpha e^2/\varepsilon l_B - \Gamma$. Here, e is the electron charge, $l_B$ is the magnetic length, $\varepsilon$ is the dielectric constant of GaAs, $\Gamma$ is the disorder broadening, and $\alpha$ is a variable. The finite layer-thickness [25,34] and Landau level mixing effects [23,35,36] were not taken into account. The optimum parameters turn out to be $\alpha$=0.00426 and $\Gamma$ = 0.23, with $\alpha$ being quite similar to the one we obtained previously [20]. Fitting according to a spin-unpolarized ground state model (i.e., $\Delta = \alpha e^2/\varepsilon l_B - g^*\mu_B B - \Gamma$, where the effective g-factor for GaAs $g^*$ = 0.44 and $\mu_B$ is the Bohr magneton) yields much worse result. Indeed, the reduced $\chi^2$ obtained from the two fittings differs by almost a factor of 10 -- 0.00004 in spin-polarized fitting versus 0.0003 in spin-unpolarized fitting. This then suggests that our density dependent result of the 5/2 energy gap support a spin polarized 5/2 state.

In summary, we have examined the impact of different kind of disorders on the experimental 5/2 energy gap. We observe that in modulation doped quantum well samples where disorder is dominated by the long-range Coulombic fluactuations the 5/2 energy gap decreases quickly with increasing $1/\mu$ or disorder. On the other hand, in HIGFETs, where the disorder is dominantly due to short-range surface roughness fluctuations, the 5/2 energy gap shows a weak mobility dependence. Moreover, our density dependent result of the 5/2 energy gap is consistent with a spin polarized 5/2 state and deviate considerable from a description in terms of a spin-unpolarized state.

We thank Dr. Z.G. Ge for his assistance at the beginning stage of this project. We would like to thank Drs. Muraki and Gamez for discussions. This work was supported by Division of Materials Sciences and Engineering, Office of Basic Energy Sciences, U.S. Department of Energy and by a Laboratory Directed Research and Development fund at Sandia. Sandia National Laboratories is a multi-program laboratory managed and operated by Sandia Corporation, a wholly owned subsidiary of Lockheed Martin company, for the U.S. Department of Energy's National Nuclear Security Administration under contract DE-AC04-94AL85000. Work at Princeton was supported on the DOE Grant No. DE-FG-02- 98ER45683. A portion of this work was carried out at the high B/T facility of the National High Magnetic Field Laboratory, which is supported by NSF Cooperative Agreement No. DMR-0084173, by the State of Florida, and by the DOE.

Figures:

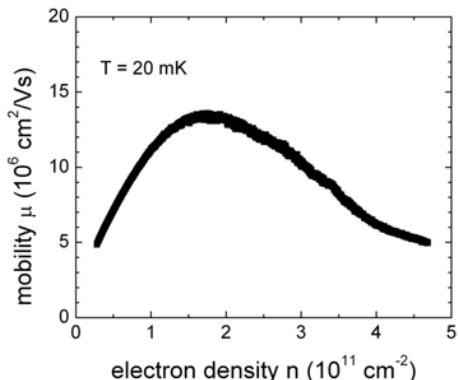

Figure 1: Electron mobility versus density in HIGFET A.

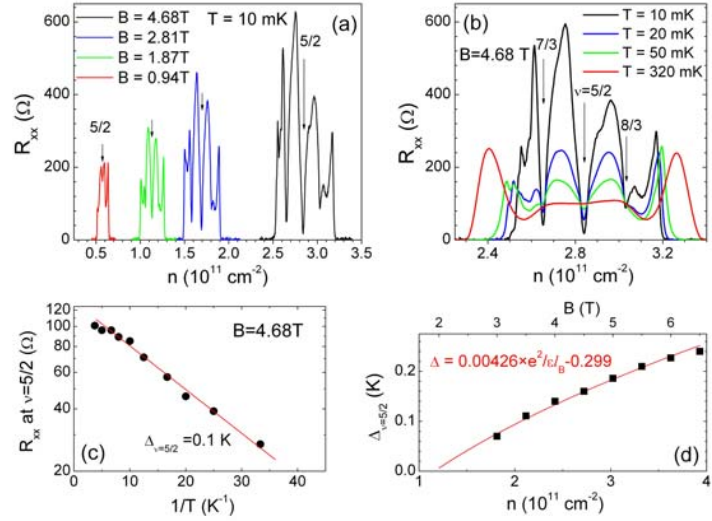

Figure 2: FQHE states in diagonal magneto-resistance around $\nu=5/2$ in HIGFET B. (a) $R_{xx}$ versus n at various magnetic fields. (b) Temperature dependence at B = 4.68T. (c) Arrhenius plot for the $R_{xx}$ minimum at $\nu=5/2$ at B = 4.68T. The line is a linear fit to the data. An energy gap of ~ 0.1K is obtained from the slope of this linear fitting. (d) Density dependence of the 5/2 energy gap in HIGFET A. The bottom x-axis is electron density, the top magnetic field. The red line is a fit according to a spin polarized model.

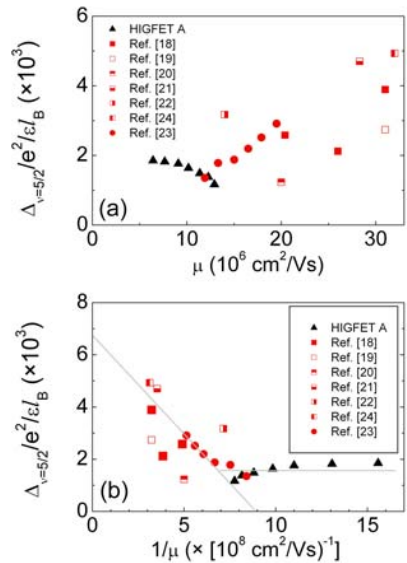

Figure 3: (a) normalized 5/2 energy gap, $\Delta_{\nu=5/2}/e^2/\varepsilon l_B$, as a function of $\mu$; and (b) as a function of $1/\mu$ for HIGFET A and modulation doped quantum well samples. The two lines in (b) are guides to the eye.

7